\documentclass[prb,aps]{revtex4}

\usepackage{graphicx}
\usepackage{dcolumn}
\usepackage{amsmath}

\newcommand{\figurewidth}{7.7cm}
\newcommand{\figureWidth}{9.3cm}
\begin{document}

\title[Correlations in Hot Dense Helium]{Correlations in Hot Dense Helium}

\author{Burkhard Militzer}

\affiliation{University of California, Berkeley, Departments of Earth and Planetary Science and Astronomy,
Berkeley, CA 94720, USA}


\begin{abstract}
Hot dense helium is studied with first-principles computer
simulations.  By combining path integral Monte Carlo and density
functional molecular dynamics, a large temperature and density
interval ranging from 1000 to 1$\,$000$\,$000 K and 0.4 to 5.4
g$\,$cm$^{-3}$ becomes accessible to first-principles simulations and
the changes in the structure of dense hot fluids can be investigated. The
focus of this article are pair correlation functions between nuclei,
between electrons, and between electrons and nuclei. The density and
temperature dependence of these correlation functions is analyzed in
order to describe the structure of the dense fluid helium at extreme conditions.
\end{abstract}


\date{\today }

\maketitle

\section{Introduction}

In the interiors of solar and extrasolar~\cite{Bu05} giant planets,
light elements such as hydrogen and helium are exposed to extreme
temperature ($\sim$10000K) and pressure (10 -- 1000 GPa) conditions.
Shock wave measurements~\cite{Ze66} are the preferred experimental
technique to reach such conditions in the
laboratory. Lasers~\cite{eggert08}, magnetic fields~\cite{Kn03}, and
explosives~\cite{Fortov2007} have recently been used to generate shock
waves that reached megabar pressures. The challenge of these
experiments is first to reach such extreme conditions and secondly to
characterize the properties of the material with sufficient accuracy
so that planetary interior model actually can be improved. Single
shock experiments yield the highest accuracy to the equation of state
(EOS). However those can rarely reach densities much higher than 4 times
the starting density, which means a large part of giant planet
interiors cannot be probed directly~\cite{Jeanloz07,MH08}.

The limitation in density has recently been addressed by combining
static compression in a diamond anvil cell with dynamic
shock compression~\cite{eggert08}. The experiments by Eggert {\it et
al.} reached pressures of 200 GPa in helium, which is a significant
increase compared to the gas gun experiments by Nellis {\it et
al.}~\cite{nellis84} from 1984 that had reached 16 GPa on the
principal Hugoniot. 

The properties of dense helium have also be the subject of a series of
recent theoretical investigations with first-principles simulation
techniques. In~\cite{Mi06}, helium was predicted to reach 5.24-fold
compression in shock experiments. The compression ratio is larger than
4 because of electronic excitations that occur in the fluid at high
temperature. Kietzmann {\it et al.}~\cite{Kietzmann07} studied the
rise in electrical conductivity using the Kubo-Greenwood formula and
compared with results of shock-wave experiments by Ternovoi {\it et
al.}~\cite{Ternovoi04}. Kowalski {\it et al.}~\cite{kowalski07}
studied dense helium in order to characterize the atmosphere of white
dwarfs. The paper went beyond the generalized gradient approximation
by considering hybrid functionals. Stixrude and
Jeanloz~\cite{StixrudeJeanloz08} studied the band gap closure in the
dense fluid helium over a wide range of densities including conditions
of giant planet interiors. Two recent studies of Jupiter's
interior~\cite{MHVTB,NHKFRB}, to very different extent, relied on a
helium EOS derived the DFT-MD.

The insulator-to-metal transition in solid helium at high pressure was
the subject of a recent quantum Monte Carlo study~\cite{KM08} that
showed that standard density functional methods underestimate the band
gap by 4 eV, which means the metallization pressure is underestimated
by 40\%.

In this article, path integral Monte Carlo (PIMC) and density
functional molecular dynamics (DFT-MD) simulations are combined to
study pair correlation functions in fluid helium over a large
density and temperature interval. This paper expands upon an earlier
work~\cite{Mi08} that demonstrated that DFT-MD results at lower
temperatures and PIMC data at higher temperatures can be combined into
one coherent EOS table. Ref.~\cite{Mi08} also provide a free energy
fit, computed adiabats and the electronic density of states. The EOS
was compared with different semi-analytical free energy
models~\cite{FKE1992,SC95,Chen2007}.

\section{Methods}

Path integral Monte Carlo~\cite{Ce95} is the most appropriate and
efficient first-principles simulation techniques for quantum system
with thermal excitations. Electrons and nuclei are treated equally as
paths, although the zero-point motion of the nuclei as well as
exchange effects are negligible for the temperatures under
consideration. The Coulomb interaction between electrons and nuclei is
introduced using pair density matrices that we derived using the
eigenstates of the two-body Coulomb problem~\cite{Po88}. The periodic
images were treated using an optimized Ewald break-up~\cite{Na95} that
we applied to the pair action~\cite{MG06}. The explicit treatment of
electrons as paths leads to the fermion sign problem, which requires
one to introduce the only uncontrolled approximation in this method,
the fixed node approximation~\cite{Ce91,Ce96}. We use the nodes from
the free-particle density matrix and from a variational density
matrix~\cite{MP00}. Besides this approximation, all correlation
effects are included in PIMC, which for example leads an exact
treatment of the isolated helium atom. We performed PIMC simulations
with 32 and 57 atoms. Additional details are given in
reference~\cite{Mi08}.

The DFT-MD simulations were performed with either the CPMD
code~\cite{CPMD} using local Troullier-Martins norm-conserving
pseudopotentials~\cite{TM91} or with the Vienna ab initio simulation
package~\cite{VASP} using the projector augmented-wave
method~\cite{PAW}. The nuclei were propagated using Born-Oppenheimer
molecular dynamics with forces derived from either the electronic
ground state or by including thermally excited electronic states when
needed. Exchange-correlation effects were described by the
Perdew-Burke-Ernzerhof generalized gradient
approximation~\cite{PBE}. The electronic wavefunctions were expanded
in a plane-wave basis with energy cut-off of 30-50 Hartrees. Most
simulations were performed with $N$=64 using $\Gamma$ point sampling
of the Brillioun zone. 

\section{Results}

\begin{figure}[!]
\centerline{\includegraphics[angle=0,width=\figureWidth]{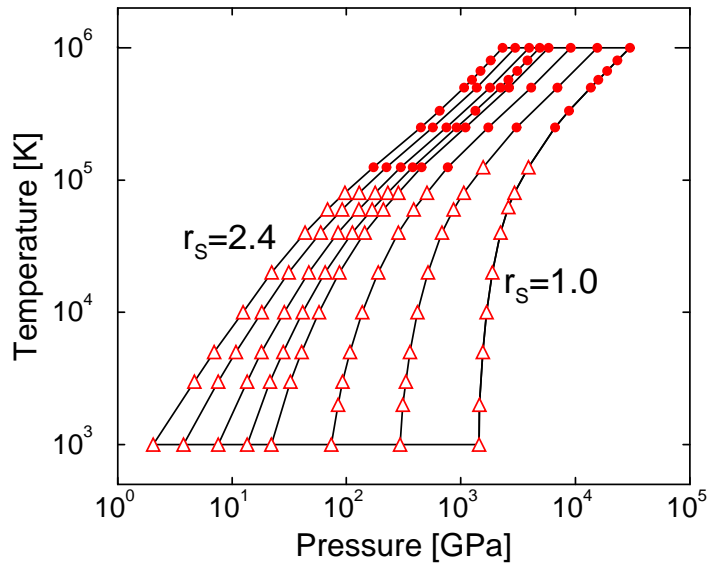}}
\caption{ Pressure-temperature diagram that indicates the conditions where 
        PIMC (circles) and DFT-MD (triangles) simulations have been
        performed. The vertical lines show isochores for the following
        $r_s$ parameters: 2.4, 2.2, 2.0, 1.86, 1.75, 1.5, 1.25, and
        1.0. }
\label{PT}
\end{figure}

Figure~\ref{PT} shows the pressure-temperature-density conditions
where PIMC and DFT-MD simulations have been preformed. The density
will be discussed in terms of the Wigner-Seitz radius, $r_s$, that is
defined by $V/N_e=\frac{4}{3}\pi (r_s a_0)^3$. There is a substantial
difference in the slopes of the isochores in figure~\ref{PT}. At low
temperature and high density, the isochores are nearly vertical
because the dominant contribution to the pressure is provided by
degenerate electrons. It takes a significant increase in temperature
before thermal excitations of the electrons add in a
noticible way to the pressure and for isochores to bend over to high
pressures. At lower densities, the biggest contribution comes from the
ionic motion, and one finds the typical pressure-temperature relation
of a dense, but not degenerate fluid.

\begin{figure}[!]
\centerline{\includegraphics[angle=0,width=\figureWidth]{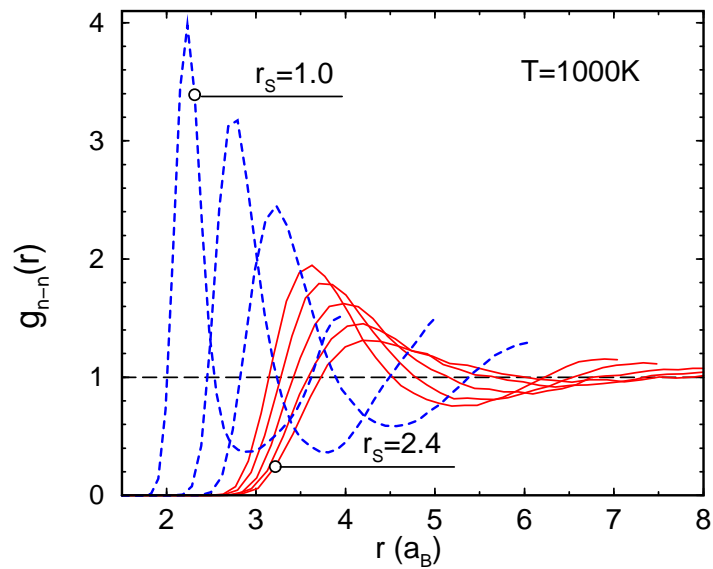}}
\caption{ 
       Nuclear pair correlation functions are shown for the 8
       different densities listed in figure~\ref{PT}.  All results
       were obtained with DFT-MD simulations at a relatively low
       temperature of 1000K. The system at the the three highest
       density (dashed lines) froze into a solid during the course of
       the MD simulation. All simulations at lower densities (solid lines) remained in a liquid state.}
\label{gnn1}
\end{figure}

Figure~\ref{gnn1} shows the nuclear pair correlation function, $g(r)$, derived
from DFT-MD simulations,
\begin{eqnarray}
g({\bf r}) &=& ~~~\frac{V}{N(N-1)} ~~\:\left< \sum_{i \ne j} \delta({\bf r}-({\bf r_i}-{\bf r_j}))\right>\;\\
g(r) &=& \frac{V}{4\pi r^2 N(N-1)}\left< \sum_{i \ne j} \delta(r-|  {\bf r_i}-{\bf r_j} |) \right>\;,
\end{eqnarray}
The sum includes all pairs of $N$ particles in volume $V=L^3$. By
definition, $g(r)$ approached 1 for large $r$ in an infinite
system. All simulations are performed in periodic boundary conditions
and we only show results for $r \le \frac{L}{2}$.

1000$\,$K is the lowest temperature under consideration. The simulations
at the highest three densities froze during
the course of the MD run. The simulations were started from scaled
fluid configurations taken from simulations at lower densities.

Figure~\ref{gnn1} shows a gradual decrease in the $g(r)$ oscillations
with decreasing density. Such oscillations are the typical signature
of the structure liquid or a solid. At high density, the $g(r)$
functions of both phases are more structured because the motion of the
particles is more confined. The $g(r)$ functions are often fairly
insensitive to the melting transition. The structure factor $S(k)$ and
the diffusion constant are more reliable measures to identify melting
in a simulation. However, that does not address the problem of
superheating and supercooling that occurs in many simulations and also
in nature. For an accurate determination of the melting temperature
one needs to perform free energy calculations or perfrom simulations
with liquid and solid in coexistence.

\begin{figure}[!]
\centerline{\includegraphics[angle=0,width=\figurewidth]{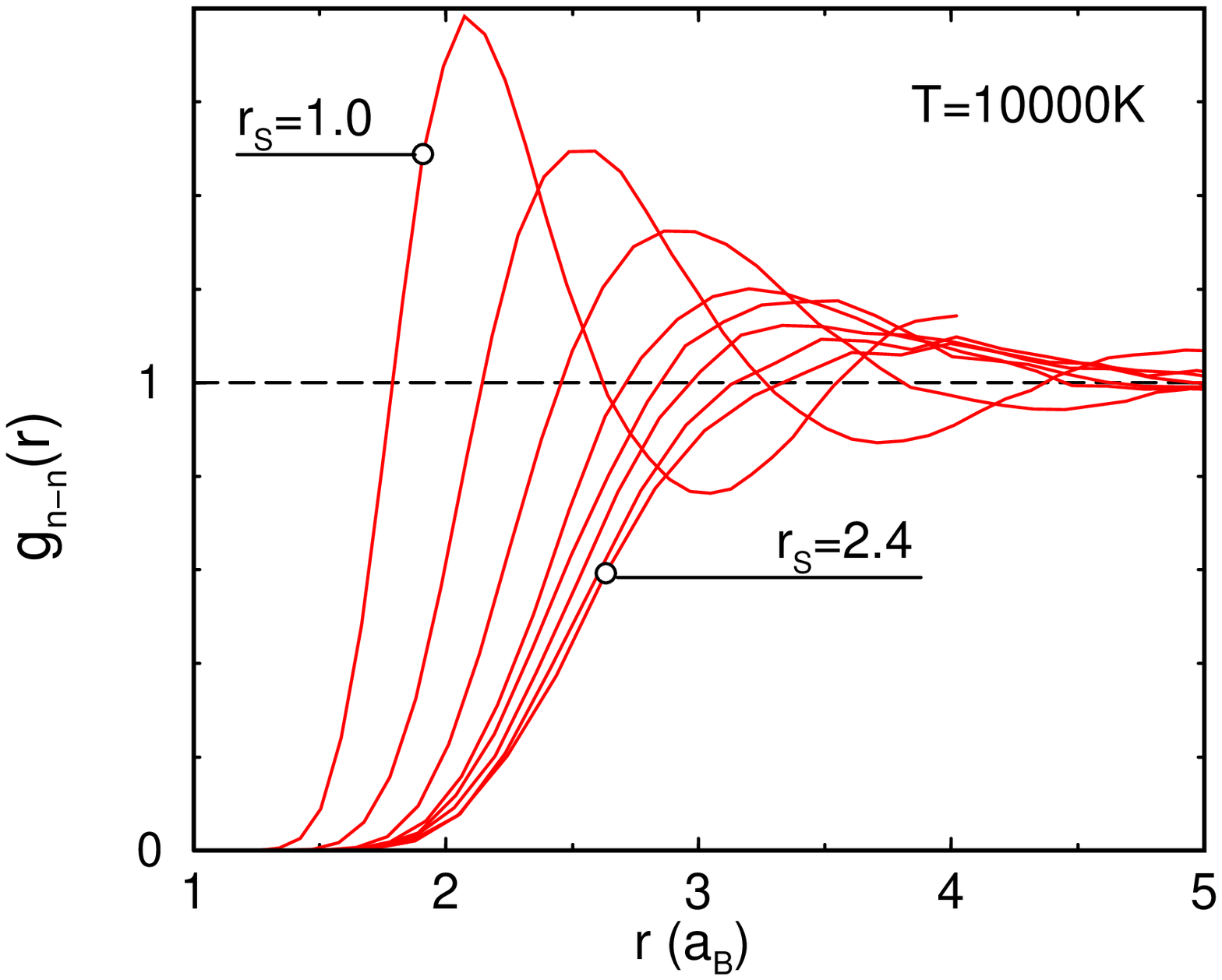}
          \includegraphics[angle=0,width=\figurewidth]{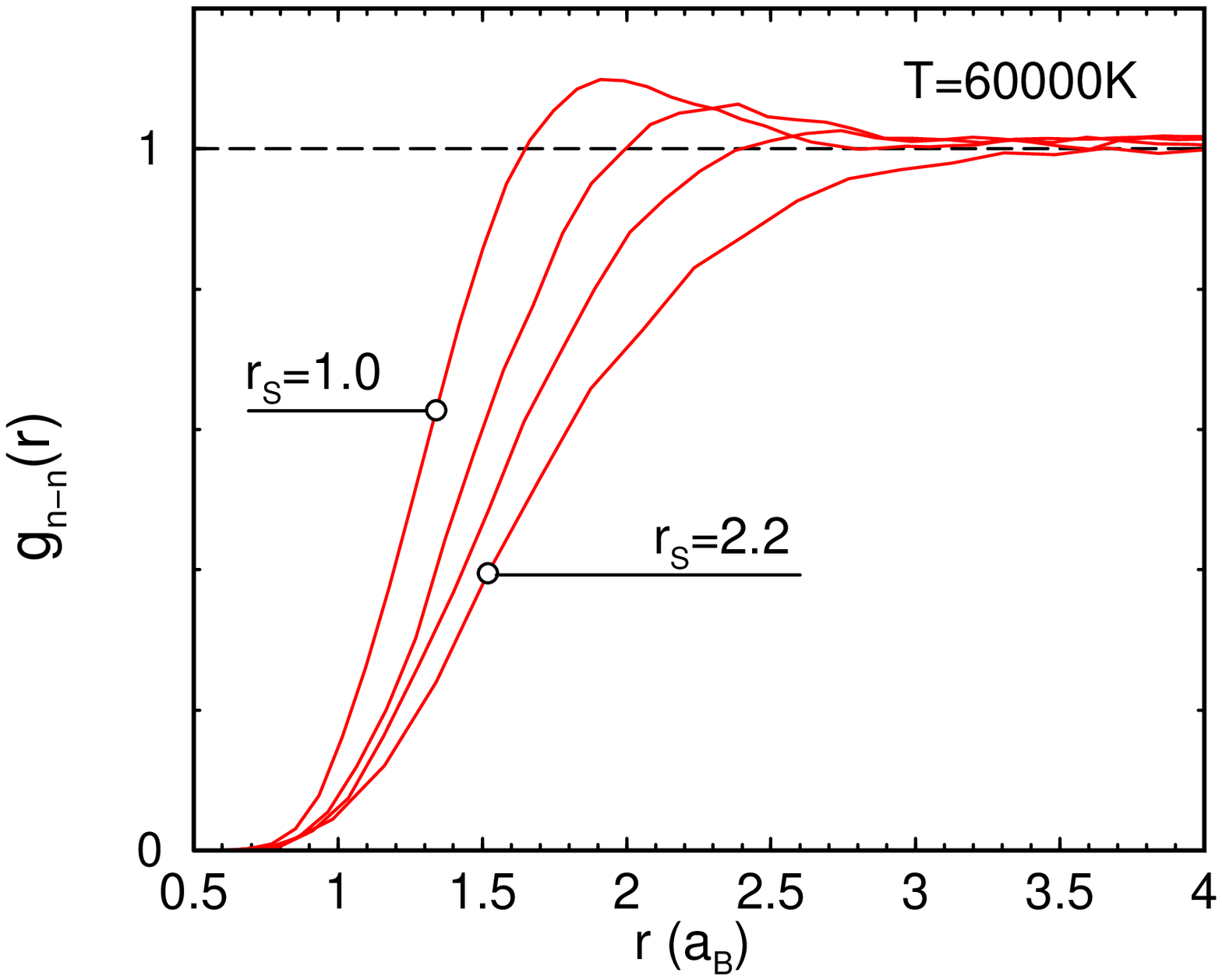}}
\caption{ 
       Nuclear pair correlation functions computed with DFT-MD are
       compared for 10$\,$000 and 60$\,$000$\,$K.  The left figure
       includes results for all densities shown in figure~\ref{PT}.
       For the right figure, only results for $r_s$=1.0, 1.25, 1.5,
       and 2.2 are shown for clarity.}
\label{gnn2}
\end{figure}

Figure~\ref{gnn2} compares the nuclear pair correlation functions at
two temperature of 10$\,$000 and 60$\,$000$\,$K. The increase in
temperature leads to stronger collisions, which means the onset of the
$g(r)$ is shifted to lower distances compared to 1000$\,$K. The motion
of particles is less confined and, consequently, the peak height in
the $g(r)$ is reduced substationally. At 60$\,$000$\,$K, the $g(r)$
shows only a small peak at the highest density, which then disappears
altogether with decreasing density.

\begin{figure}[!]
\centerline{\includegraphics[angle=0,width=\figurewidth]{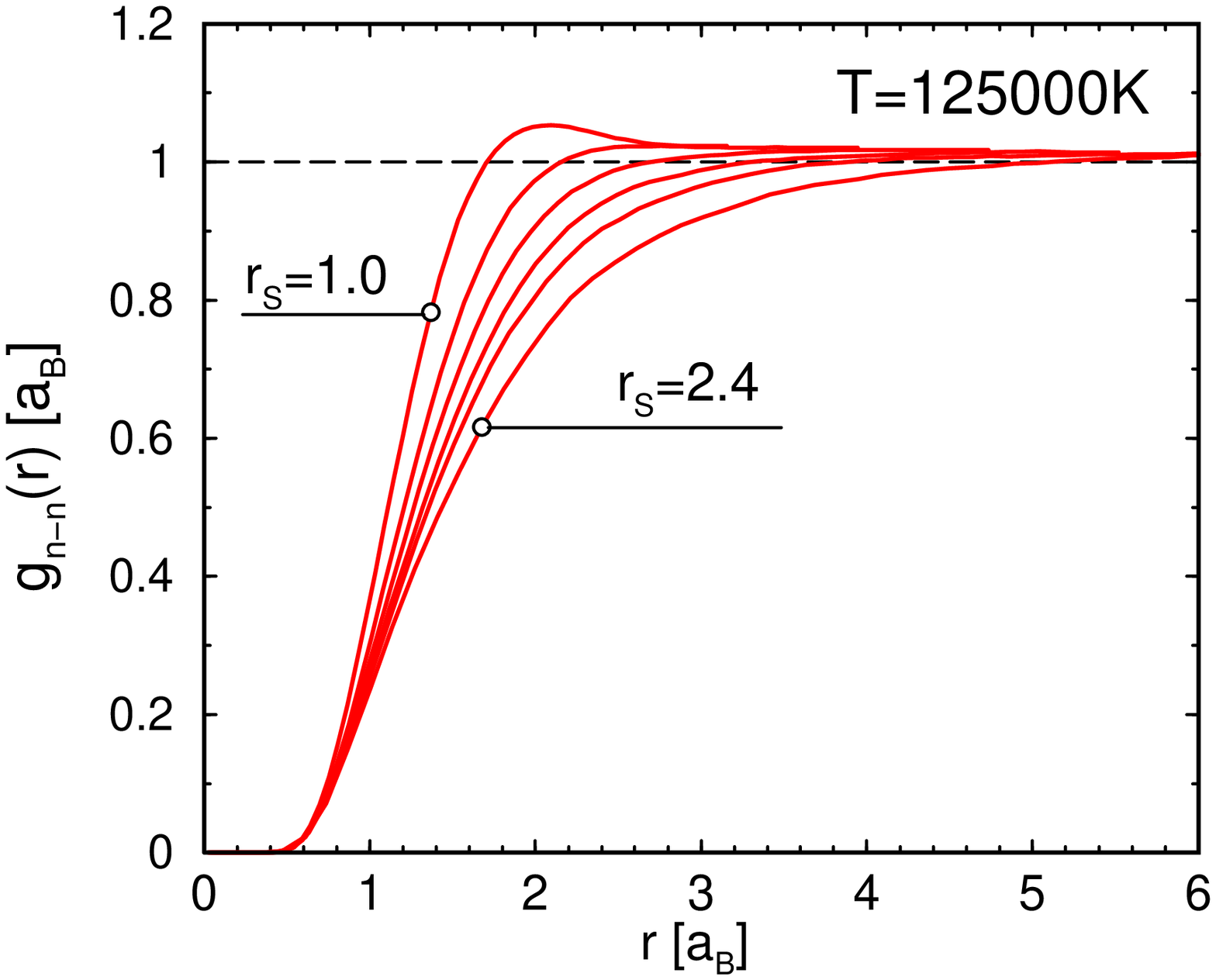}
          \includegraphics[angle=0,width=\figurewidth]{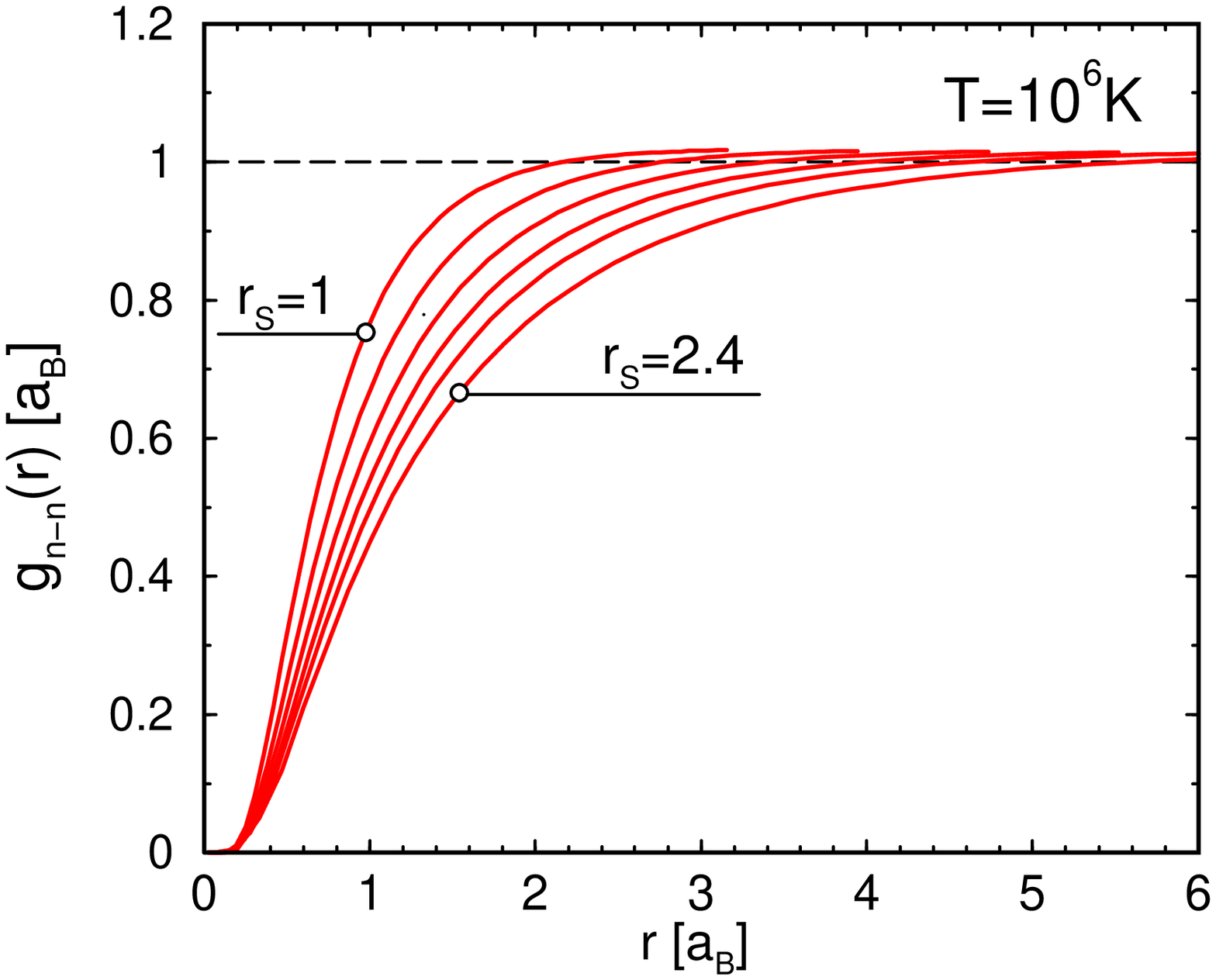}}
\caption{ 
       Nuclear pair correlation functions computed with PIMC are shown for 
       $125\,000$ (left) and $10^6\,$K (right) are shown for the following $r_s$ values: 1.0,
       1.25, 1.5, 1.75, 2.0, and 2.4. }
\label{gnn3}
\end{figure}

Figure~\ref{gnn3} shows the nuclear pair correlation function derived
with PIMC for 125$\,$000 and 10$^6\,$K. The expected trend continues;
with increasing temperature and decreasing density the fluid becomes
less structured. At an extreme temperature of 10$^6\,$K, one can
hardly find any positive correlation in the motion of the nuclei for
the whole density range under consideration.

\begin{figure}[!]
\centerline{\includegraphics[angle=0,width=\figurewidth]{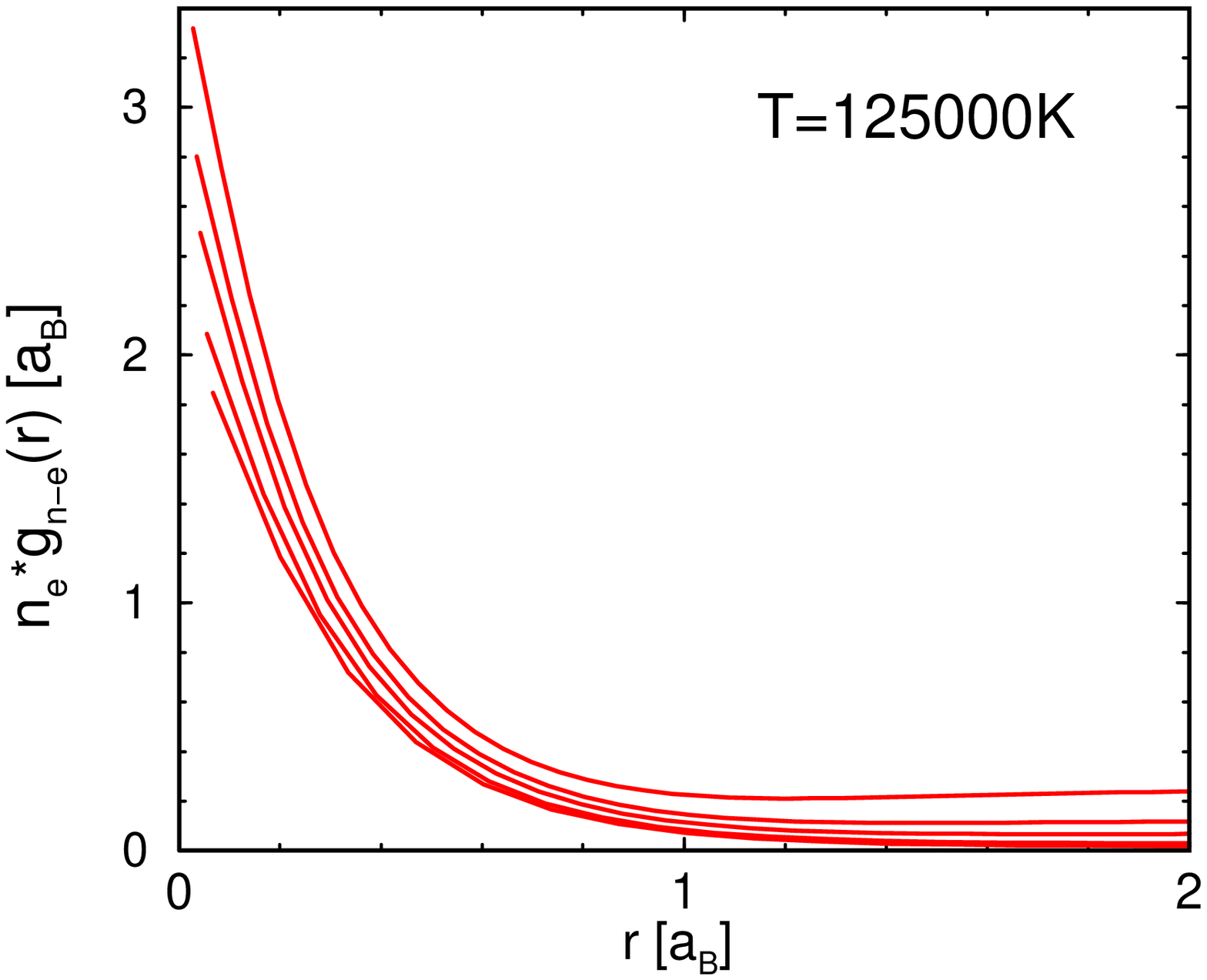}
          \includegraphics[angle=0,width=\figurewidth]{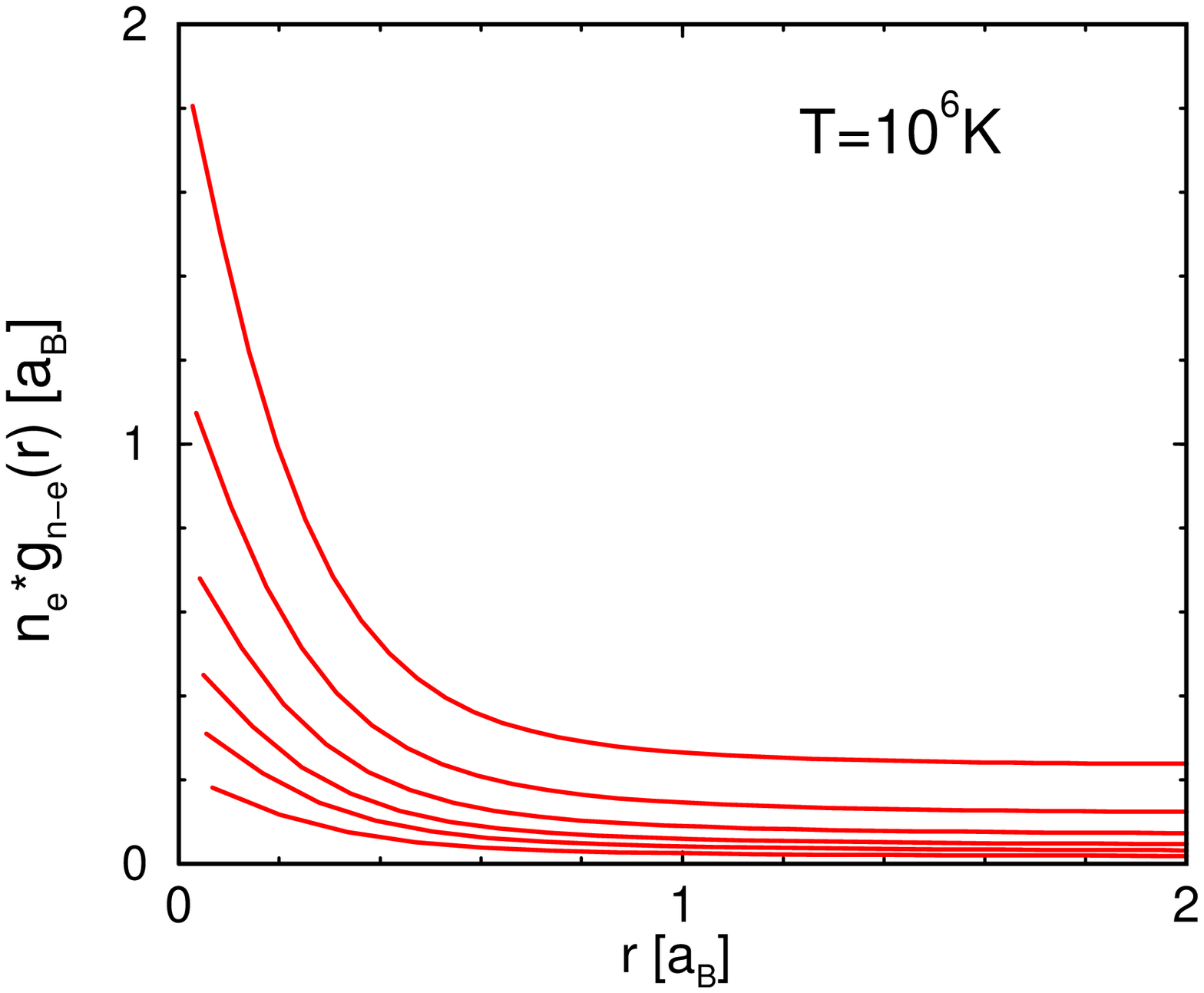}}
\caption{ 
       Electron-nucleus pair correlation functions, $g_{n-e}(r)$, are
       shown for $125\,000\,$ (left) and 10$^6$K (right) for the
       following $r_s$ values: 1.0 (top line), 1.25, 1.5, 1.75 (only
       for 10$^6$K) 2.0, and 2.4 (lowest line). The $g_{n-e}(r)$
       function has been multilied by the volume density of electrons
       so that the area under the curves is related to the fraction of
       bound electrons. }
\label{g_ne}
\end{figure}

Figure~\ref{g_ne} shows the correlation functions between nuclei and
electrons. The $g_{n-e}(r)$ functions were multiplied by the electronic
density, $N_e/V$, so that integral under the curve is related the
number of electrons within a certain distance from a nucleus. This
provides a qualitative estimate for the fraction of bound electrons. A
quantititive estimate is difficult to give because unbound electrons
in Rydberg scattering states also make contributions to $g_{n-e}(r)$ at small
$r$. Furthermore, there is no straightforward way to tell to which
nucleus a particular electron is bound. The long tail in the
$g_{n-e}(r)$ includes unbound electrons as well as contribution from
electrons bound to other nuclei nearby. In principle, the
fraction of bound electrons can be determined from the natural orbitals of
the reduced single particle density matrix~\cite{Gi90}, but no
practical method that is compatible with many-body simulations has been
advanced. At very low density, however, approaches that fit $g(r)$
have worked reasonably well~\cite{Mi98}.

At 125$\,$000$\,$K, the electron-nucleus pair correlation functions in
figure~\ref{g_ne} show a relatively weak dependence on density. The
presence of the peak at small $r$ suggests that a fraction of the
electrons are still bound at this temperature. At 10$^6\,$K, the
density dependence of the $g_{n-e}(r)$ is much stronger. At low
density, there hardly any positive correlation in the motion of
electrons and nuclei, which indicates the nuclei are almost
fully ionized and most electrons are free. At higher densities, the
motion of the electrons is more confined, which leads to an increased positive
correlation with the nuclei.

\begin{figure}[!]
\centerline{\includegraphics[angle=0,width=\figurewidth]{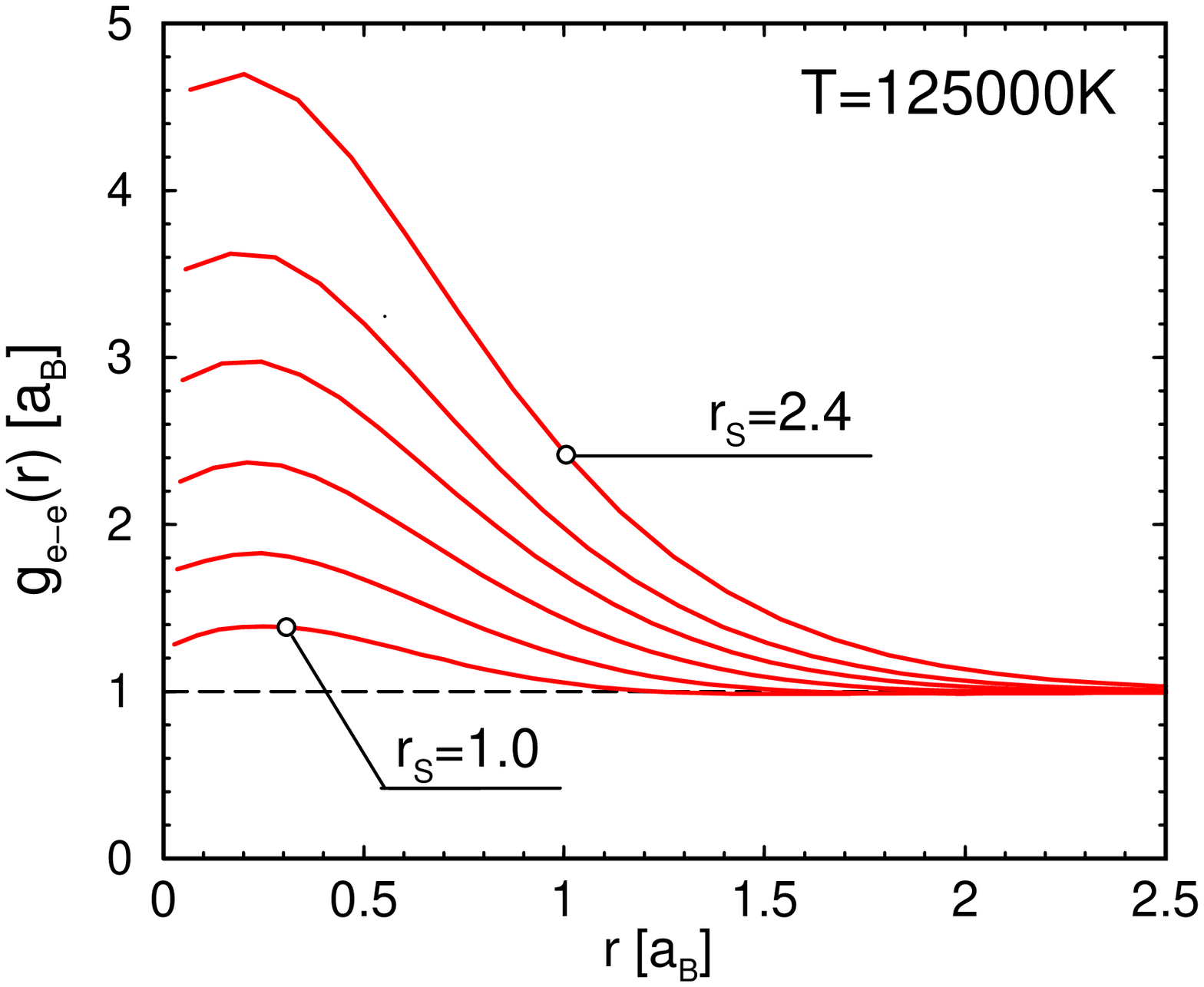}
          \includegraphics[angle=0,width=8cm]{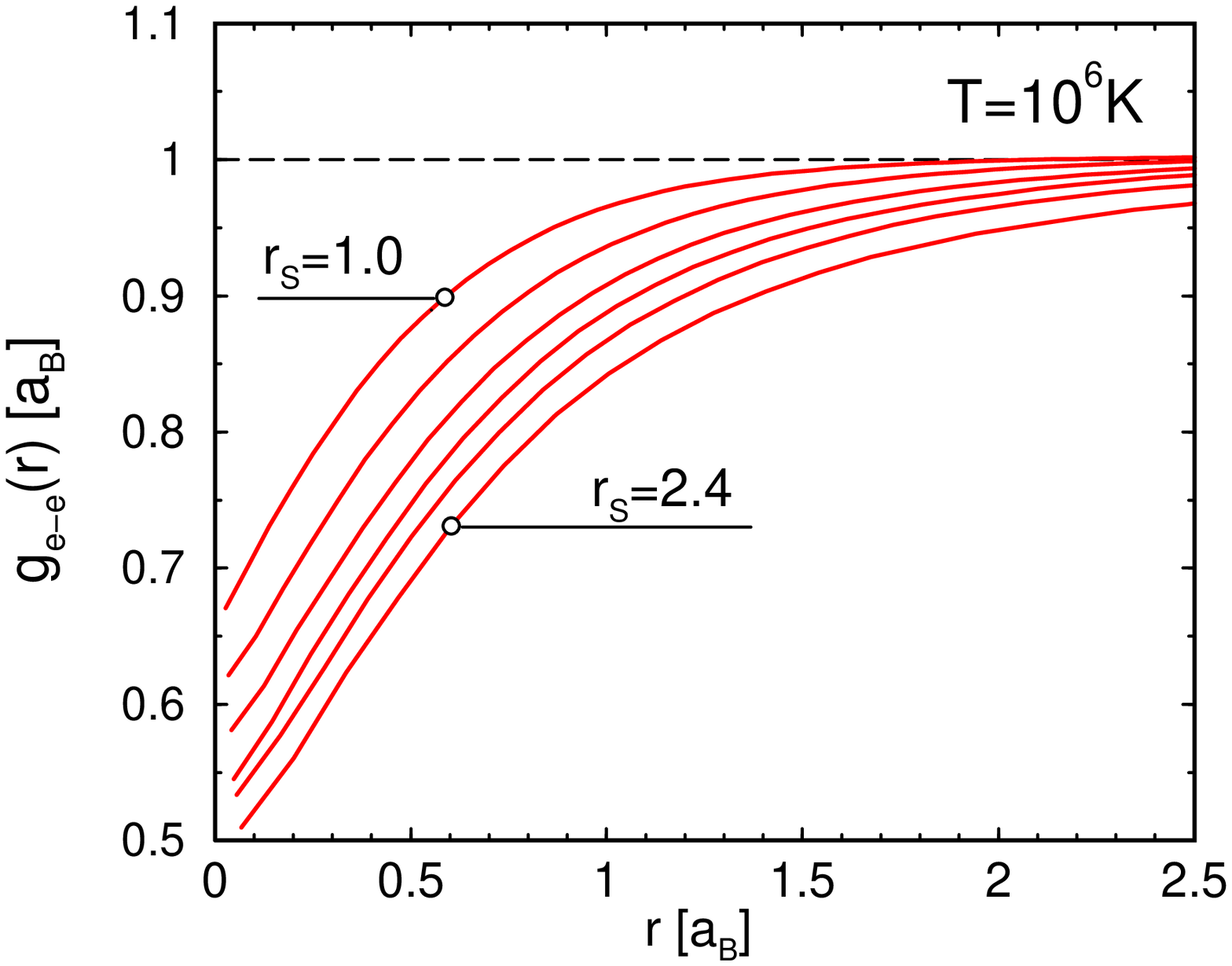}}
\caption{ 
       The electron-electron pair correlation functions for electrons
       with opposite spins are shown for 125$\,$000$\,$K on the left
       and for 10$^6\,$K on the right.  The lines correspond to
       different densities represented by the following $r_s$
       parameters: 1.0, 1.25, 1.5, 1.75, 2.0, and 2.4.}
\label{gr_ee_u}
\end{figure}

The pair correlation function of the electrons with opposite spin in
figure~\ref{gr_ee_u} shows a qualitatively different behavior at low
and high temperature. Since they have different spins, the Pauli
exclusion principle does not apply and only Coulomb repulsion keeps
these pairs of  particles apart. However, at 125$\,$000$\,$K, one observe a
significant positive correlation, because two electrons with opposite
spins are bound in a helium atom. So the peak in this $g_{e-e}(r)$
confirms the existance of bound electrons. The peak height decreases
significantly with increasing density, which suggests that degeneracy
in the electron gas increases and the number of bound states decreases
accordingly. The slope of the isochores in figure~\ref{PT} underlines
the importance of the degeneracy at 125$\,$000$\,$K.

At 10$^6$ K the behavior of the pair correlation function of the
electrons with opposite spin is very different. The electrons are
mostly unbound and are not degenerate. Coulomb repulsion leads to a
negative correlation that much weaker than the Pauli exclusion for
electron with parallel spin to be discussed next. With increasing
density, the Coulomb repulsion has less of an effect, which means the
value of $g(r=0)$ increases.

\begin{figure}[!]
\centerline{\includegraphics[angle=0,width=\figurewidth]{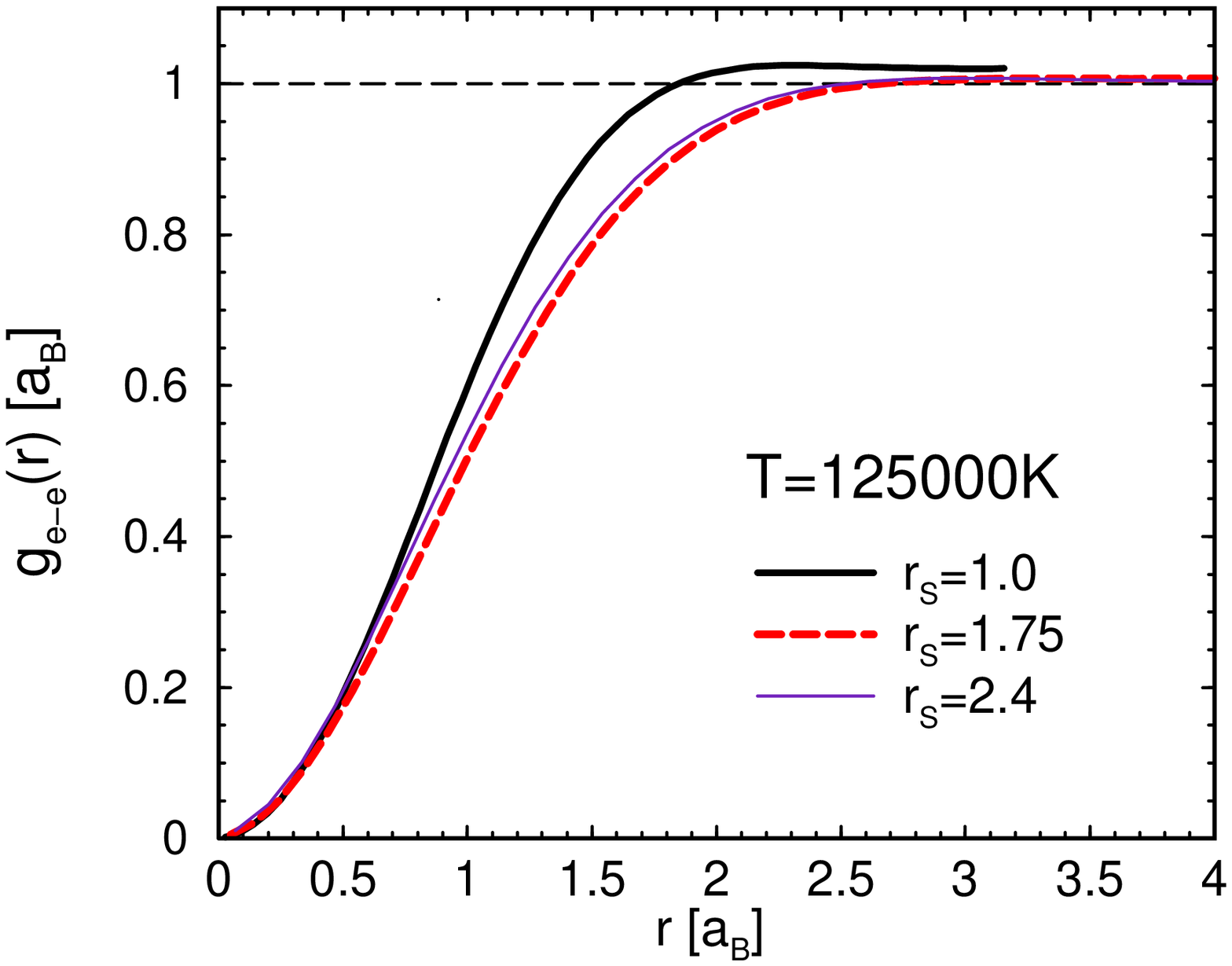}
\includegraphics[angle=0,width=7.85cm]{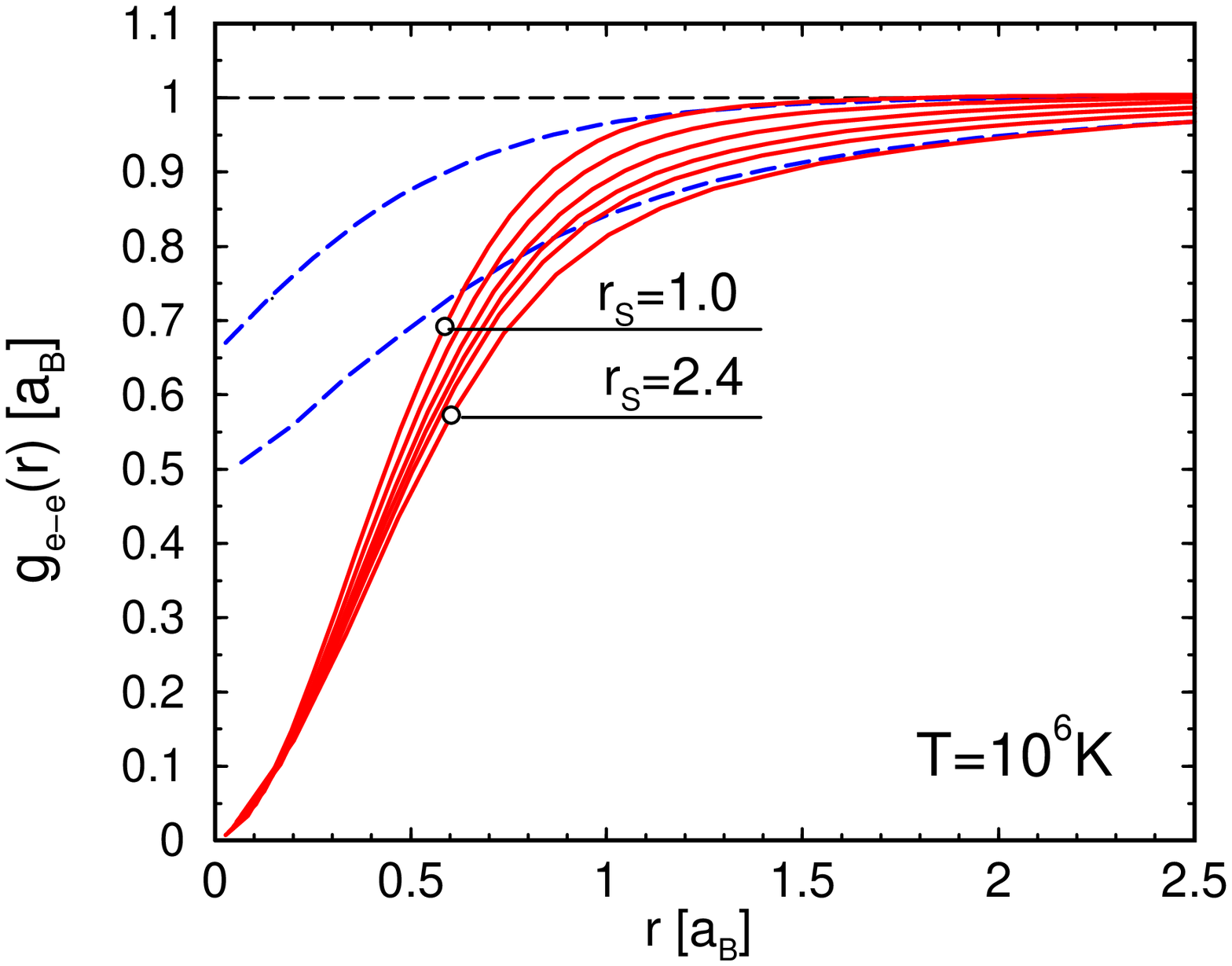}}
\caption{ 
       The electron-electron pair correlation functions for electrons
       with parallel spins are shown for 125$\,$000$\,$K on the left,
       and as solid lines (see figure~\ref{gr_ee_u} for $r_s$ values)
       for 10$^6\,$K on the right. The dashed lines on the right
       repeat correlation functions for electrons with opposite spins
       from figure~\ref{gr_ee_u} for $r_s=1.0$ and 2.4.}
\label{gr_ee_l}
\end{figure}

Figure~\ref{gr_ee_l} shows the pair correlation function of the
electrons with parallel spins. The negative correlation at small $r$
is known as exchange-correlation hole. The effect of the Pauli
exclusion is short-ranged, which can be understood from the graph at
10$^6\,$K. At approximately $r=1.3\,a_0$, the correlation functions of
electrons with parallel and opposite spin converge, which marks the
point where Pauli exclusion is no longer relevant.

The density dependence of the $g(r)$ of electrons with parallel spins
at 125$\,$000$\,$K is more difficult to interpret because of the
presence of bound states. At $r_s=1$, one finds a small positive
correlation, which can only be explained with a positive correlation
in the nuclei $g(r)$ (figure~\ref{gnn3}) and the presence of bound
electrons. Furthermore, one expects the size of the
exchange-correlation hole would increase with decreasing density,
which explains the difference between $r_s=1$ and 1.75. However, if
the density is decreased further the size of the exchange-correlation
hole shrinks slightly. Since there are more bound electrons present at
$r_s=2.4$, the collision of the nuclei brings the electron closer
together than they would be otherwise, which explains observed behavior.

In conclusion, the structure of a very simple fluid, helium, has been
discussed at extreme pressure and density conditions. The observed
trends are general and have relevance to other light elements are
conditions of planetary interiors.

\acknowledgements
A part of the reported first-principles simulations were performed at NERSC and at NCSA.

\section*{References}


\end{document}